\documentclass[fleqn,10pt]{wlscirep}
\PassOptionsToPackage{bookmarks=false}{hyperref}
\usepackage[utf8]{inputenc}
\usepackage[T1]{fontenc}

\title{Influence of intermixing at the Ta/CoFeB interface on spin Hall angle in Ta/CoFeB/MgO heterostructures}

\author[1,*]{Monika Cecot}
\author[2]{\L{}ukasz Karwacki}
\author[1]{Witold Skowro\'{n}ski}
\author[1]{Jaros\l{}aw Kanak}
\author[3]{Jerzy Wrona}
\author[4]{Antoni \.Zywczak}
\author[5] {Lide Yao}
\author[5] {Sebastiaan van Dijken}
\author[2, 6]{J\'{o}zef Barna\'{s}}
\author[1]{Tomasz Stobiecki}

\affil[1]{AGH University of Science and Technology, Department of Electronics,  Al. Mickiewicza 30, 30-059 Krak\'{o}w, Poland}
\affil[2]{Faculty of Physics, Adam Mickiewicz University, ul. Umultowska 85, 61-614 Pozna\'{n}, Poland}
\affil[3]{Singulus Technologies AG, Kahl am Main, 63796, Germany}
\affil[4]{Academic Center of Materials and Nanotechnology, AGH University of Science and Technology, Al. Mickiewicza 30, 30-059 Krak\'{o}w, Poland}
\affil[5] {NanoSpin, Department of Applied Physics, Aalto University School of Science, P.O. Box 15100,
FI-00076 Aalto, Finland}
\affil[6]{Institute of Molecular Physics, Polish Academy of Sciences, ul. Smoluchowskiego 17, 60-179 Pozna\'{n}, Poland}

\begin{abstract}
We investigate the spin Hall effect in perpendicularly magnetized Ta/Co$_{40}$Fe$_{40}$B$_{20}$/MgO trilayers with Ta underlayers thicker than the spin diffusion length. The crystallographic structures of the Ta layer and Ta/CoFeB interface are examined in detail using X-ray diffraction and transmission electron microscopy. The thinnest Ta underlayer is amorphous, whereas for thicker Ta layers a disoriented tetragonal $\beta$-phase appears. Effective spin-orbit torques are calculated based on harmonic Hall voltage measurements performed in a temperature range between 15 and 300 K. To account for the temperature dependence of damping-like and field-like torques, we extend the spin diffusion model by including an additional contribution from the Ta/CoFeB interface. Based on this approach, the temperature dependence of the spin Hall angle in the Ta underlayer and at Ta/CoFeB interface are determined separately. The results indicate an almost temperature-independent spin Hall angle of $\theta_{SH}^{N}\approx -0.2$  in Ta and a strongly temperature-dependent $\theta_{SH}^{I}$ for the intermixed Ta/CoFeB interface.
\end{abstract}
\begin{document}

\flushbottom
\maketitle
%
%
\thispagestyle{empty}

\section*{Introduction}

It is well known that spin current induced {\it via} the spin Hall effect (SHE) in a heavy metallic layer may exert a torque on the magnetic moment of an adjacent magnetic layer\cite{hoffmann2013spin}. This torque in turn can induce magnetization dynamics, which may be observed by various experimental techniques, namely spin-torque ferromagnetic resonance \cite{liu2011spin}, spin-orbit-torque-induced magnetization switching \cite{hao2015giant}, domain-wall movement \cite{emori2013current} or harmonic Hall voltage measurements \cite{kim2013chirality}.
Recent reports indicate that signals detected in the ferromagnetic layer do not refer directly to the spin Hall angle, considered as an {\it intrinsic} property of the nonmagnetic layer and defined as a ratio of the induced spin current density to the charge current density in this layer~\cite{liu2011spin}.
Consequently, additional torques due to spin-orbit interaction at the interface between the nonmagnetic and magnetic layers are invoked. This interfacial spin-orbit interaction leads to a nonequilibrium spin polarization of electrons at the interface, which in turn gives rise to spin-orbit torque (SOT). There is currently a great interest, both experimental and theoretical, in the interfacial SOTs. However, it is rather difficult experimentally to distinguish between the interface SOT and the torque induced by spin current generated in the nonmagnetic layer by SHE and injected to the magnetic layer. Theoretical approaches to the SOT usually assume a sharp interface\cite{haney2013current,wang2016giant}. Such an approximation can be justified in some cases, but it is generally invalid in amorphous structures. For example, in CoFeB/Ta and CoFeB/W\cite{skowronski_W_2016} structures, X-ray and neutron reflectometry\cite{zhu2012study} prove strong intermixing between the two layers, which results in a relatively wide interfacial region.
When considering the interface contribution to the total spin torque, one needs to take into account spin transport across the interfaces and possible mechanisms of spin relaxation. Accordingly, one introduces an efficiency parameter, or equivalently an effective spin Hall angle. Since the SOT consists of two components -- damping-like and field-like torques, two spin Hall torque efficiencies have been proposed \cite{pai2015dependence}.

Investigation of the impact of interface atomic ordering and electrical conductivity on the spin torque arising from SHE is crucial for the design of novel spintronic devices utilizing SOT. High resistance phases of heavy metals are desirable as the spin Hall angle is enhanced, e.g. tungsten above a certain thickness shows a phase transition from the high resistivity $\beta$-W phase to the low resistivity $\alpha$-W phase, which results in a decrease of the spin Hall angle \cite{pai2012spin, hao2015beta, skowronski_W_2016}. 
The whole situation seems to be rather complex due to a relatively broad distribution of experimental data obtained on samples from different laboratories and not unified definitions, which do not allow to draw definite conclusions. For instance, the reported values of the spin Hall angle in systems with Ta span between -0.03 and -0.15 (at room temperature)~\cite{zhang2013magnetotransport, allen2015experimental, kim2014anomalous, qiu2014angular, liu2012spin}. Moreover, one should make a distinction between heterostructures with in-plane and perpendicular magnetic anisotropy (PMA) of the ferromagnetic (FM) layer. In the case of structures with PMA (like those considered in this paper), harmonic Hall voltage method~\cite{pi2010tilting, hayashi2014quantitative, akyol2015effect} seems to be the most appropriate because the damping-like and field-like torques can be determined directly from independent measurements.

In the case studied in this paper, i.e. in Ta/CoFeB bilayers, no transition between $\alpha$-Ta and $\beta$-Ta phase is observed, and the growth of a given phase is determined exclusively by sputtering conditions. For measurements we selected a CoFeB layer with PMA and Ta buffer layer thicker than the spin diffusion length, which falls within the range 1.2-2.5 nm~\cite{kwon2015influence, behera2016anomalous, allen2015experimental, morota2011indication}. Since the available experimental data on the damping-like torque, converted to effective spin Hall angle, fluctuate between -0.03 \cite{zhang2013magnetotransport} and -0.11 \cite{qiu2014angular}, we find it desirable to investigate the effects due interface and crystallographic structure, which may be responsible for the scattering of reported data.
We provide characterization of the microstructure data on electrical conductivity and show that both depend on the Ta underlayer thickness, beyond the range already investigated \cite{ou2016strong, kim2016spin}. The thinnest Ta layer is amorphous, whereas thicker Ta layers comprise the tetragonal $\beta$ phase. In all cases, the interface effects originate from mixing of Ta and CoFeB,which results in a relatively thick {\it interface layer}.

Structures of micrometer-dimensions are investigated with harmonic Hall voltage measurements in a wide range of temperatures in order to clearly designate the field-like and damping-like torques. Because of substantial interlayer mixing, we propose to model transport properties by considering the interface as a distinct layer with its own spin diffusion length and spin Hall angle. 

\section*{Results}

\begin{figure}[!ht]
\centering
\includegraphics[width=4.0in]{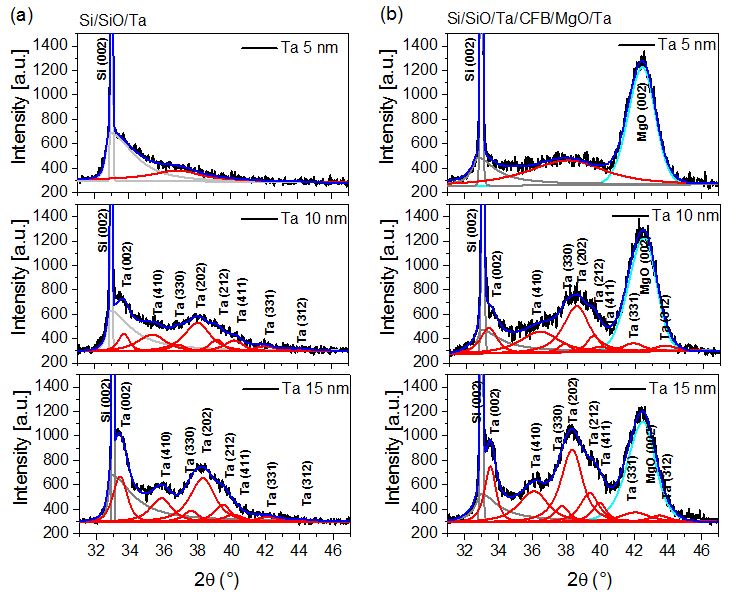}
\caption{$\theta$$-2$$\theta$ profiles for single Ta layers (a), and for full multilayer stacks Ta($d_{\rm Ta}$/Co$_{40}$Fe$_{40}$B$_{20}(1)$/MgO(5)/Ta(3) (b), where d$_{Ta}$ = 5, 10, 15 nm. Black line is the experimental data, blue line is a fitting, red lines represent distribution of the Ta peaks for different orientations, cyan line is a fitting for MgO, gray lines show substrate peaks.}
\label{fig:XRD}
\end{figure}

\begin{figure}[!ht]
\centering
\includegraphics[width=3.0 in]{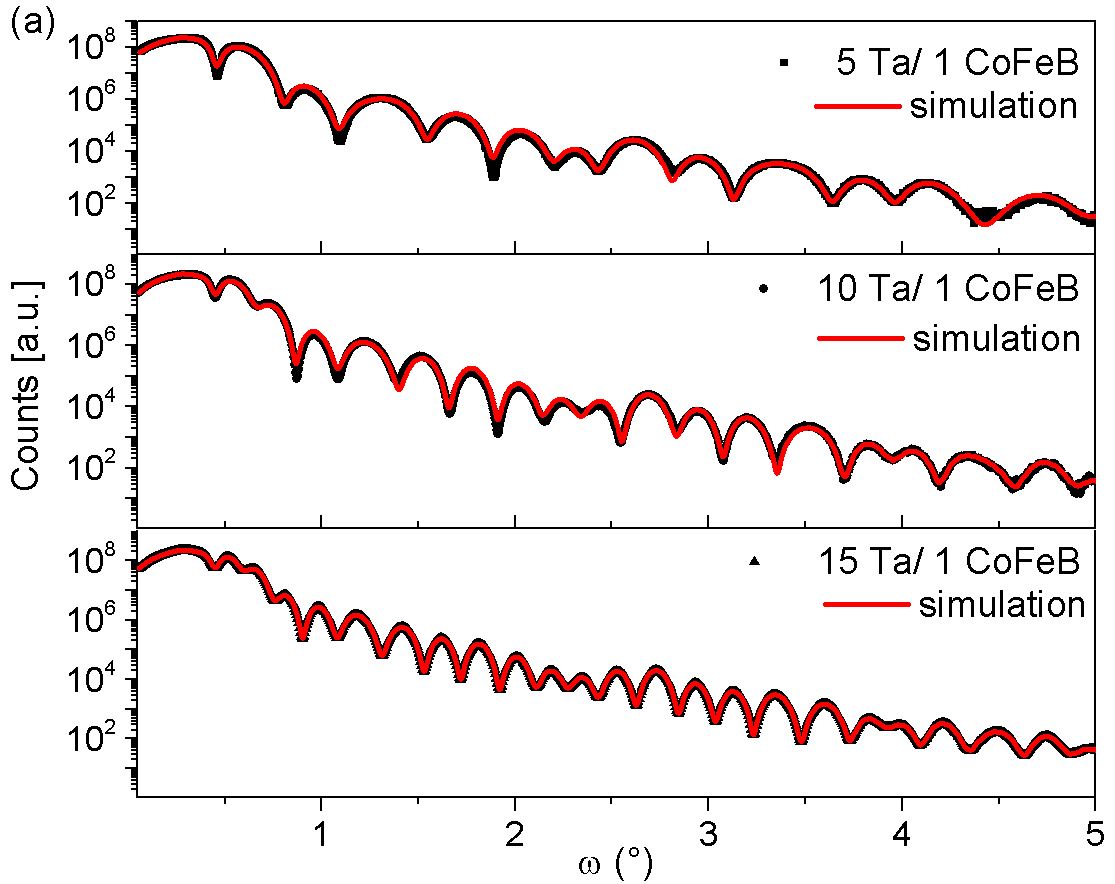}
\includegraphics[width=1.1 in]{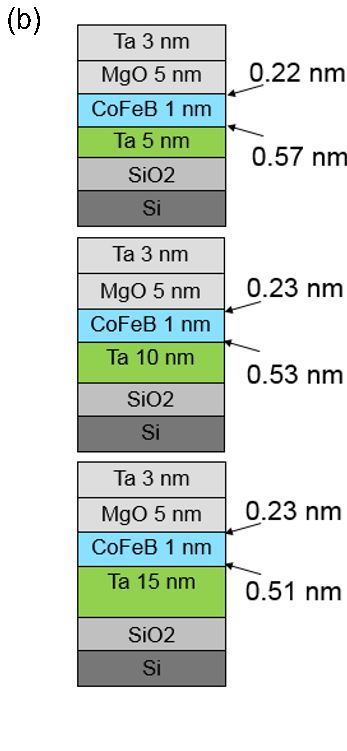}
\caption {The best fits to x-ray reflectivity data for full structures: Ta(d$_{\rm Ta}$)/CoFeB(1)/MgO(5)/Ta(3), where d$_{\rm Ta}$ = 5, 10, 15 nm (a), and interfacial RMS roughness for Ta/CoFeB and CoFeB/MgO interfaces (b), determined with an accuracy of $\pm0.02$ nm.}
\label{fig:XRR2}
\end{figure}

Two sets of samples were prepared. The first set consists of Ta($d_{\rm Ta}$)/Co$_{40}$Fe$_{40}$B$_{20}$/MgO(5)/Ta(3) multilayer stacks, with d$_{Ta}$ = 5, 10, 15 and $t_{FM}$ = 0.8-1.5, while structures in the second set comprised of single Ta layers of thickness d$_{Ta}$ (all thicknesses in nm). Structural analysis of the Ta layer was performed on single Ta layers with $d_{Ta}$ = 5, 10, 15 nm (Fig. \ref{fig:XRD}(a)) and on annealed Ta($d_{\rm Ta})$/Co$_{40}$Fe$_{40}$B$_{20}$(1)/MgO(5)/Ta(3) multilayer stacks, with the identical Ta thickness (Fig. \ref{fig:XRD}(b)). Comparisons between the $\theta$$-2$$\theta$ X-ray Diffraction (XRD) profiles of Ta in the full stacks and in single layers do not reveal major structural differences. The spectra of 5 nm thick Ta layers show a very broad low-intensity reflection indicating an amorphous-like disordered structure. On the other hand, spectra of thicker Ta layers (10 nm and 15 nm) contain peaks that can be attributed to a polycrystalline tetragonal $\beta$ phase.
The $\theta$$-2$$\theta$ XRD measurements on these samples do not indicate the presence of the $\alpha$-Ta phase. For all samples, the MgO layers have a highly (001)-oriented texture, while the thin CoFeB layers remain amorphous after annealing.
Subsequently, using the X-ray Reflectivity (XRR) method we analysed Ta/CoFeB and CoFeB/MgO interfaces and corresponding spectra are presented in Fig. \ref{fig:XRR2} (a). The thickness of the CoFeB/MgO interface is about 0.23 nm, while that of the Ta/CoFeB interface is in the range from 0.51 nm to 0.57 nm, see Fig. \ref{fig:XRR2} (b). The RMS roughness from Atomic Force Microscope (AFM), measured on the surface of single Ta layers, is comparable to the thickness of CoFeB/MgO interface from XRR. The RMS roughness are as follows: 0.23 nm, 0.26 nm and 0.29 nm for 5 nm, 10 nm and 15 nm of Ta, respectively. As expected, the smoothest surface is for the amorphous 5 nm Ta sample \cite{liu2015correlation}, while the polycrystalline Ta layers (10 and 15 nm) are increasingly rough. A significant difference between the widths of the interfaces Ta/CoFeB and CoFeB/MgO can be explained by mechanisms described below. Small thickness of the interface CoFeB/MgO mainly comes from topological roughness, what indicates poor interdiffusion between the CoFeB and MgO layers. In turn, large thickness of the Ta/CoFeB interface may have an origin in a large negative interfacial enthalpy, which is a driving force for interdiffusion. It is worth noting that for Fe in Ta the interfacial enthalpy is -54 kJ/(mole of atoms) and for Co in Ta it is -86 kJ/(mole of atoms)\cite{de1988cohesion}. In addition, the XRR measurements results implicate a tendency for the interface thickness to diminish with increasing Ta layer thickness. This phenomena can be explained by easier interdiffusion when both CoFeB and Ta (5 nm) are amorphous than between amorphous CoFeB and polycrystalline Ta (10 nm and 15 nm) layers. The described tendency is in accordance with decrease in thickness of the magnetic dead layer (MDL), see Table. 1.

\begin{figure}[!ht]
\centering
\includegraphics[width=4.5 in]{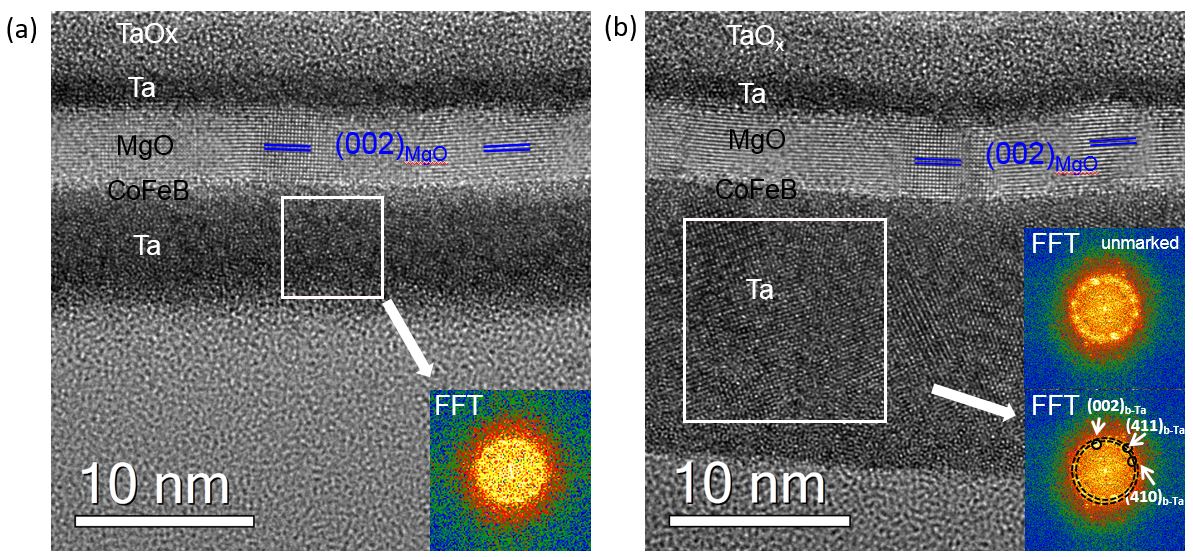}
\caption {HRTEM images of Ta($d_{Ta}$)/CoFeB(1)/MgO(5)/Ta(3) samples. (a) $d_{Ta}$ = 5 nm; (b) $d_{Ta}$ = 15 nm. The (002) planes in the MgO layers are marked with blue lines. The insets in (a) and (b) show fast Fourier transform (FFT) patterns from designated areas. The results indicate that the 5 nm Ta layer is amorphous and the 15 nm Ta layer is polycrystalline.}
\label{fig:HRTEM}
\end{figure}

High-resolution Transmission Electron Microscope (TEM) images of samples with 5 nm Ta and 15 nm Ta are shown in Fig. \ref{fig:HRTEM}. The layer thicknesses of each multilayer stack correspond closely to the intended growth parameters. In both samples, the MgO layer exhibits a polycrystalline structure with the (002)-planes oriented parallel to the interfaces, while the CoFeB layer is amorphous. The main difference between the two samples is the crystalline structure of the Ta underlayer. The 5 nm thick Ta layer is amorphous (Fig. \ref{fig:HRTEM}(a)), whereas the 15 nm Ta layer is polycrystalline (Fig. \ref{fig:HRTEM}(b)).The crystallographic orientations in the 15 nm thick Ta layer are in agreement with the $\theta$$-2$$\theta$ XRD measurements.

\begin{figure}[!ht]
\centering
\includegraphics[width=4.5 in]{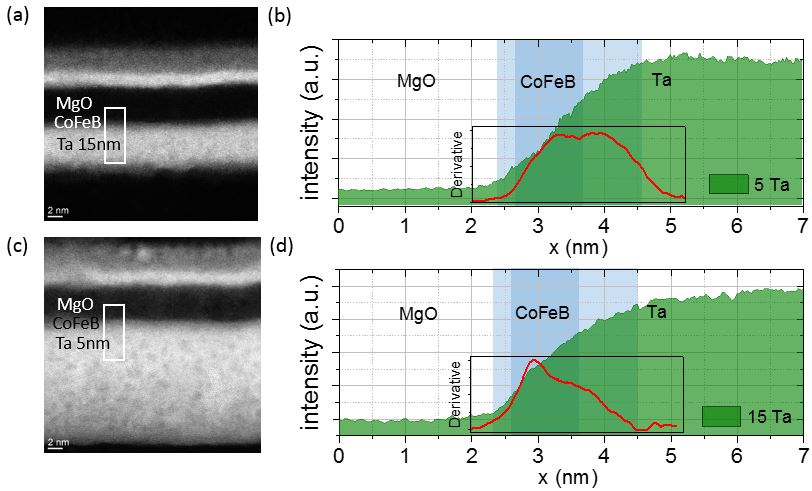}
\caption {(a),(c) Z-contrast images and (b),(d) Z-contrast profiles from the indicated areas for samples with (a),(b) 5 nm Ta and (c),(d) 15 nm Ta. The blue areas indicate the nominal CoFeB thickness and the light blue areas visualize the range of interlayer mixing. The insets show derivatives of the Z-contrast profiles.}
\label{fig:TEM}
\end{figure}

The interface between Ta and CoFeB is mixed in both samples. TEM measurements reveal a gradual change in Z-contrast, as illustrated by the line profiles of Fig. 4(b) and (d). The more gradual increase in Z-contrast in Fig. 4(d) suggests that atomic interdifussion changes slightly when the Ta layer thickness is increased from 5 nm to 15 nm. This effect is most likely caused by the crystal structure of the Ta layers, which evolves from amorphous to polycrystalline when the film becomes thicker. Variations in intermixing and the crystal structure can both affect electronic transport across the Ta/CoFeB interface.
In order to apprehend the difference between Z-contrasts we propose the following scenario.
For the sample with 5 nm Ta layer, diffusion takes place between two amorphous layers and thus the interdiffusion zone is more or less homogeneous. Accordingly, derivative of the Z-contrast contour is constant in the CoFeB area (inset to Fig.\ref{fig:TEM}(b)).
However, in the case of 15 nm Ta the diffusion process takes place between amorphous CoFeB and polycrystalline Ta layers, and the migration mechanism appears mainly along Ta grain boundaries. Therefore, shape of the the Z-contrast contour close to Ta is rounded and goes slowly into saturation (Fig.\ref{fig:TEM}(d)).

\subsection*{Temperature dependence of electrical and magnetic properties}

\begin{figure}[!ht]
\centering
\includegraphics[width=2.0in]{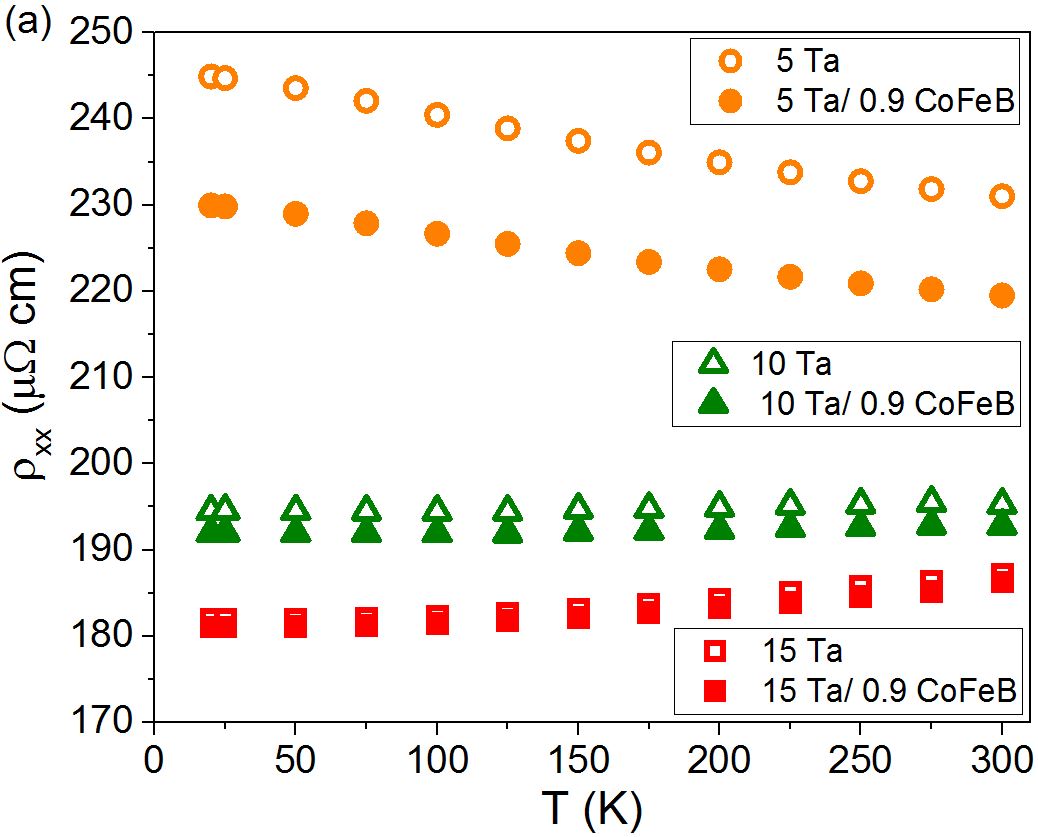}
\includegraphics[width=2.0in]{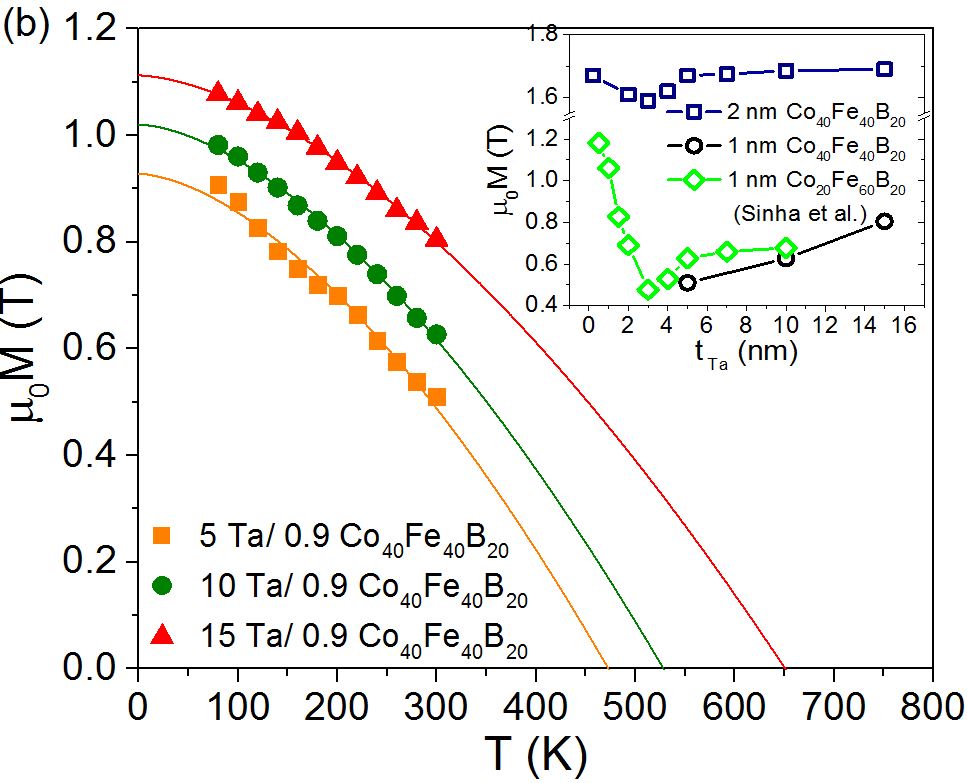}
\includegraphics[width=2.0in]{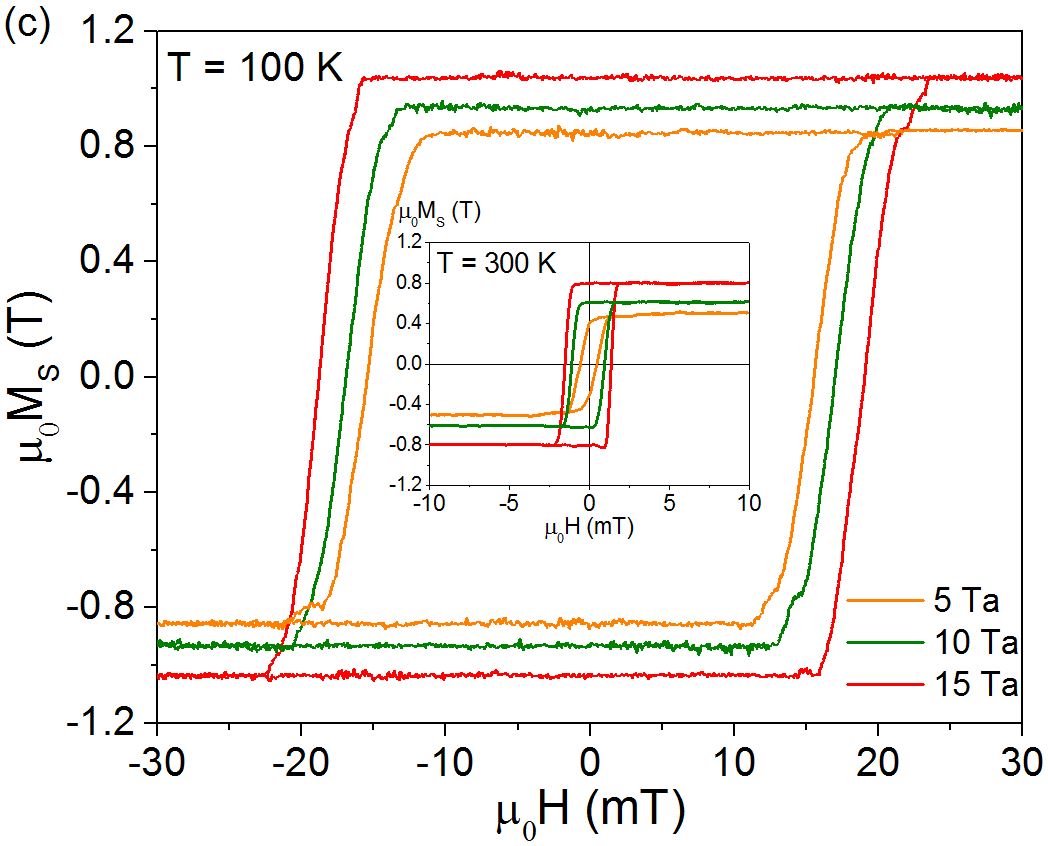}
\caption{(a) Temperature dependence of the longitudinal resistivity for the single and stack layers. (b) Spontaneous magnetization M as a function of temperature; solid lines represent the results calculated from the Bloch's law; inset: saturation magnetization M$_{S}$ of CoFeB vs. Ta underlayer thickness, comparison of our results for 1 nm and 2 nm of Co$_{40}$Fe$_{40}$B$_{20}$ with the data after Ref. \cite{sinha2013enhanced} for 1 nm Co$_{20}$Fe$_{60}$B$_{20}$. (c) Magnetization hysteresis loops obtained at 100K  and at room temperature (inset).}
\label{fig:Ms}
\end{figure}

The longitudinal resistivity $\rho_{xx}$ on the full multilayer stacks and also for the single Ta layers were measured using a 4-probe method. The room-temperature resistivities of single Ta layers are as follows: $\rho_{5 Ta}$ = 235 $\mu\Omega$cm, $\rho_{10 Ta}$ = 195 $\mu\Omega$cm, $\rho_{15 Ta}$ = 185 $\mu\Omega$cm. The resistivity of Ta can be used as a probe of  structural ordering. The high resistivity, of more than 200 $\mu\Omega$cm confirms the amorphous structure of 5 nm Ta. In turn, resistivities of the order of 190 $\mu\Omega$cm evidence the presence of the $\beta$-Ta phase in 10 and 15 nm Ta layers \cite{clevenger1992relationship, stella2009preparation}. The obtained resistivities of $\beta$-Ta are similar to those reported in Refs.\cite{allen2015experimental,avci2014fieldlike}.
The resistivity of amorphous CoFeB, $\rho_{\rm CoFeB} \approx$ 165 $\mu\Omega$cm, is concluded from the parallel resistors model. Temperature dependence of the longitudinal resistivities ($\rho_{xx}$) for the full stack as well as for single Ta layers are presented in Fig.\ref{fig:Ms} (a). Interestingly, resistivities of the full multilayer stacks samples change very little with temperature -- the highest resistivity change is 4$\%$ for the sample with the thinnest Ta layer. In this case, negative temperature coefficient of the resistivity is noticed, which is characteristic of  the amorphous phase.
For $d_{Ta}=5$ nm, resistivity of the single layer is noticeably larger than that of the corresponding stack layer. For $d_{Ta}=10$ the difference is rather small. All this support the conjecture that the layers 10 nm and 15 nm of Ta are in the $\beta$ phase, while 5 nm of Ta is amorphous.

We examined magnetization of the CoFeB layer as a function of temperature for samples with different Ta thickness (Fig. \ref{fig:Ms} (b)). The temperature dependence of the spontaneous magnetization is described by the Bloch's law:
$ M= M_{0}(1-(T/T_{c})^{3/2}$,
where $M_{0}$ is a spontaneous magnetization at T = 0K and $T_{C}$ is the Curie temperature (see Table I).
A strong changes of the CoFeB layer saturation magnetization with Ta thickness form an additional evidence of the amorphous phase and interdiffused interface for thin Ta layers. In prepared comparison of CoFeB saturation magnetization obtained at room temperature and literature reports~\cite{sinha2013enhanced, kim2014anomalous} one can note a drop in the saturation magnetization for $d_{Ta}$ $\approx$ 3 nm (see the inset to Fig. \ref{fig:Ms} (b)). This dependence is also noticed for a larger thickness of CoFeB (2 nm). It clearly shows that thin Ta amorphous layers are mixed with amorphous CoFeB at the interface, which results in a strong reduction of the saturation magnetization.

\begin{table}[!ht]
\renewcommand{\arraystretch}{1.3}
\label{tab:Static}
\centering
\begin{tabular}{c|ccccc}
$d_{Ta}$ & MDL & $\mu_{0}M_{0}$ & $T_{C}$ & $\mu_{0}M_{S}$ \\
(nm) & (nm) & (T) & (K) & (T) \\
\hline
5 & 0.55 & 0.93 & 472 & 0.50 \\
10 & 0.46 & 1.02 & 528 & 0.63 \\
15 & 0.39 & 1.11 & 650 & 0.80 \\
\end{tabular}
\caption{Thickness of magnetic dead layer (MDL) and spontaneous magnetization $\mu_0M_0$  at T=0K, Curie temperature $T_C$, and saturation magnetization $\mu_0M_S$  at 300K, for the three Ta thicknesses, $d_{Ta}$.}
\end{table}

Magnetization hysteresis loops (Fig. \ref{fig:Ms} (c)) confirmed perpendicular magnetization in the annealed samples. However, the magnetic hysteresis loop for the sample with 5 nm of Ta shows weaker perpendicular magnetic anisotropy than other samples. This is a result of transition to the superparamagnetic state, which probably appears due to local thickness discontinuity in the CoFeB layer resulting from a thick magnetic dead layer. The fact that strong  mixing at the Ta/CoFeB interface can result in a MDL, was already reported before~\cite{chen2012magnetic, sinha2013enhanced, frankowski2015buffer}. Actually, MDL for the sample with 5 nm of Ta is the widest (Table I). An increased thickness of MDL reflects interdiffusion at the Ta/CoFeB interface.

\subsection*{Anomalous Hall effect}

Anomalous Hall voltage measurements were performed for external magnetic field applied perpendicularly to the sample plane and temperature changing from 20 K to 300 K. The anomalous Hall effect (AHE) is described quantitatively by the anomalous Hall coefficient $R_s$, which can be calculated according to the relation $\rho_{AHE} = R_S \; \mu_0 M_S$, where $\rho_{AHE}$ is the anomalous Hall resistivity.

\begin{figure}[!ht]
\centering
\includegraphics[width=2.0in]{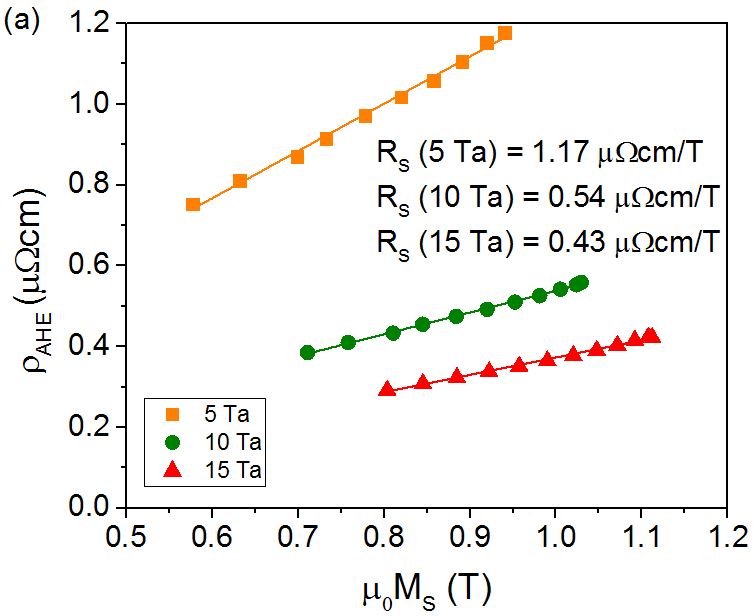}
\includegraphics[width=2.0in]{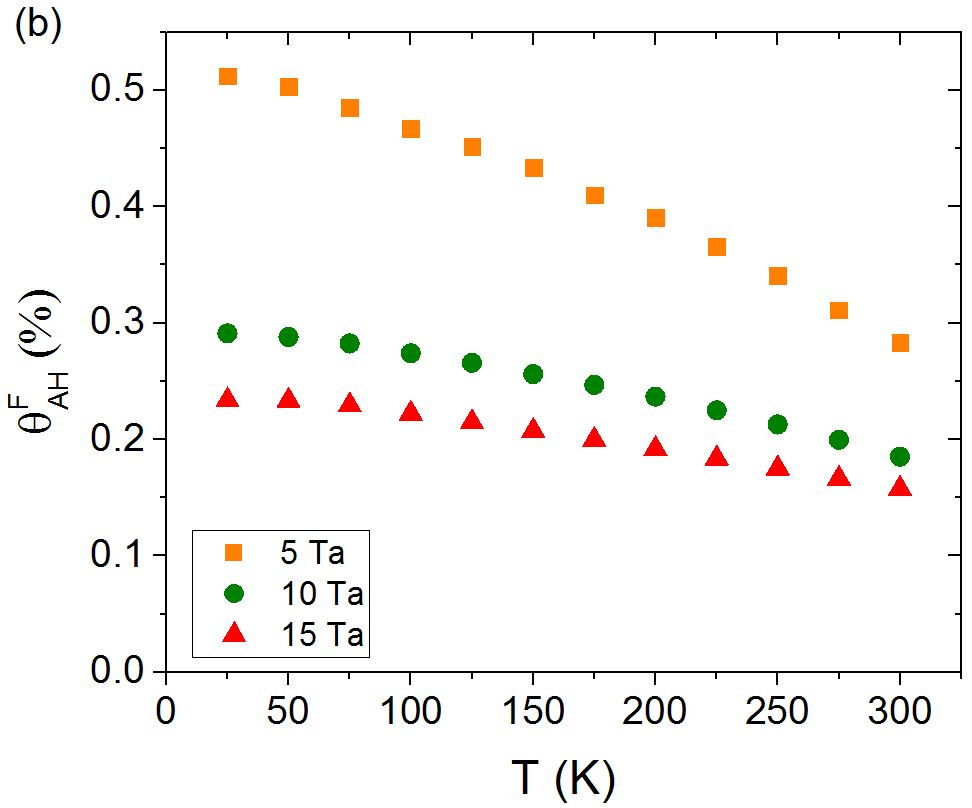}
\caption{ (a) Anomalous Hall resistivity of Ta/CoFeB/MgO as a function of saturation magnetization -- slope determines the corresponding anomalous Hall coefficient $R_{S}$. (b) Temperature dependence of the anomalous Hall angle, $\theta_{AH}^F$.}
\label{fig:AHE}
\end{figure}

Figure \ref{fig:AHE}(a) presents the anomalous Hall resistivity as a function of the saturation magnetization. The variation of $M_{S}$ is derived from temperature dependent data. Slopes of these curves determine the corresponding anomalous Hall coefficients. According to the theoretical predictions~\cite{smit1958spontaneous, berger1970side}, $R_S$ is proportional to the resistivity $\rho_{xx}$. Insignificant change of the resistivity in our samples, which is less than 4\% between 20K and 300K (Fig.\ref{fig:Ms}(a)), indicate that the anomalous Hall coefficients are roughly independent of temperature. The twice higher value of $R_S$ for the sample with 5 nm Ta in comparison to other samples points to a substantial influence of the Ta/CoFeB interface. The anomalous Hall angle, defined as the ratio of the Hall resistivity $\rho_{AHE}$ to the longitudinal resistivity $\rho_{xx}$, is  shown in Fig. \ref{fig:AHE}(b). The results are comparable to those reported in Refs~\cite{wu2013anomalous, zhu2014giant}.

The planar Hall resistance was measured in an external magnetic field applied in the film plane with rotation of the field direction. In contrast to W/CoFeB heterostructures, the planar Hall effect (PHE) for Ta underlayers is much smaller than the corresponding AHE contribution~\cite{torrejon2014interface,kim2014anomalous}.
The ratio  of planar Hall resistance, $R_{PHE}$, and the anomalous Hall resistance, $R_{AHE}$, is of the order of $\zeta = R_{PHE}/R_{AHE}\approx 0.03$, which is within the margin of error, and therefore PHE can be neglected. This significantly simplifies the formula for effective torque fields, as described below.

\subsection*{Spin torque efficiencies}

\begin{figure}[!ht]
\centering
\includegraphics[width=4.0in]{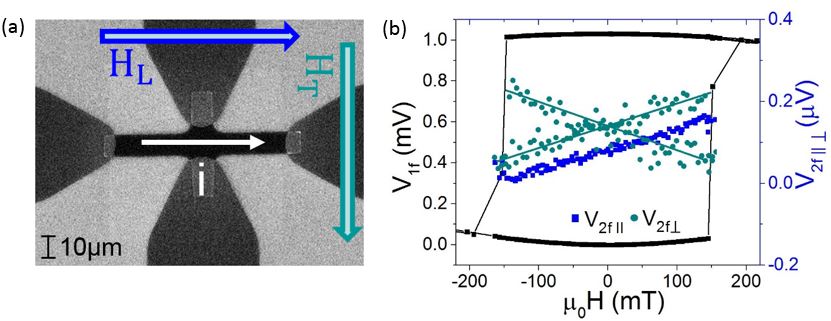}
\caption{(a) SEM image of the Hall bar with illustrated directions of current flow and external magnetic field: longitudinal H$_{L}$ and transverse H$_{T}$ to the current. (b) First harmonic (1f) and second harmonic (2f)  Hall voltage signals (longitudinal and transverse in the latter case).}
\label{fig:scheme2}
\end{figure}

To evaluate the spin torque exerted on the CoFeB layer we determined the spin torque efficiencies. Therefore, the Hall voltage measurements were performed for external magnetic field applied in the sample plane in two directions: along and transversely to the current flowing through the Hall bar. Details of the measurement technique are described e.g. in references \cite{pi2010tilting,hayashi2014quantitative,skowronski_W_2016}.
Lock-in technique was used to measure first and second harmonics of the Hall voltage. Low frequency (385 Hz) alternating charge current was passed through the Hall bar, while the external magnetic field was applied along or across the Hall bar, as indicated in Fig. \ref{fig:scheme2}(a). Measurements were performed for the Hall bars 10 $\mu$m wide and 40 $\mu$m long. In order to determine the current density, the precise widths of the patterned strips were controlled by scanning electron microscope (SEM).

From the harmonic Hall voltage measurements, the spin-orbit induced effective fields (related to the spin torques) were obtained as a function of temperature for all the three studied samples (for different values of $d_{Ta}$). From the measurements for longitudinal external magnetic field, $H_{L}$, we derived the effective field  $\Delta H_\mathrm{DL}$. Analogously, from the measurements for transverse external field, $H_{T}$, we obtained the effective field  $\Delta H_\mathrm{FL}$. Taking into account that $\zeta = R_{PHE}/R_{AHE}$ is negligibly small, these effective fields are determined by the voltage harmonics according to the formulas
\begin{equation}
\Delta H_\mathrm{DL(FL)}=-2{\dfrac{\partial V_\mathrm{2f}}{\partial H_\mathrm{L(T)}}}/{\dfrac{\partial^2 V_\mathrm{1f}}{\partial H_\mathrm{L(T)}^2}},
\label{eq:field}
\end{equation}
where $V_{1f, 2f}$ are the first and second harmonic Hall signals measured for $H_{L}$ and $H_{T}$. Exemplary voltage signals measured at  150 K are presented in Fig. \ref{fig:scheme2}(b).

\begin{figure}[!ht]
\centering
\includegraphics[width=4.0in]{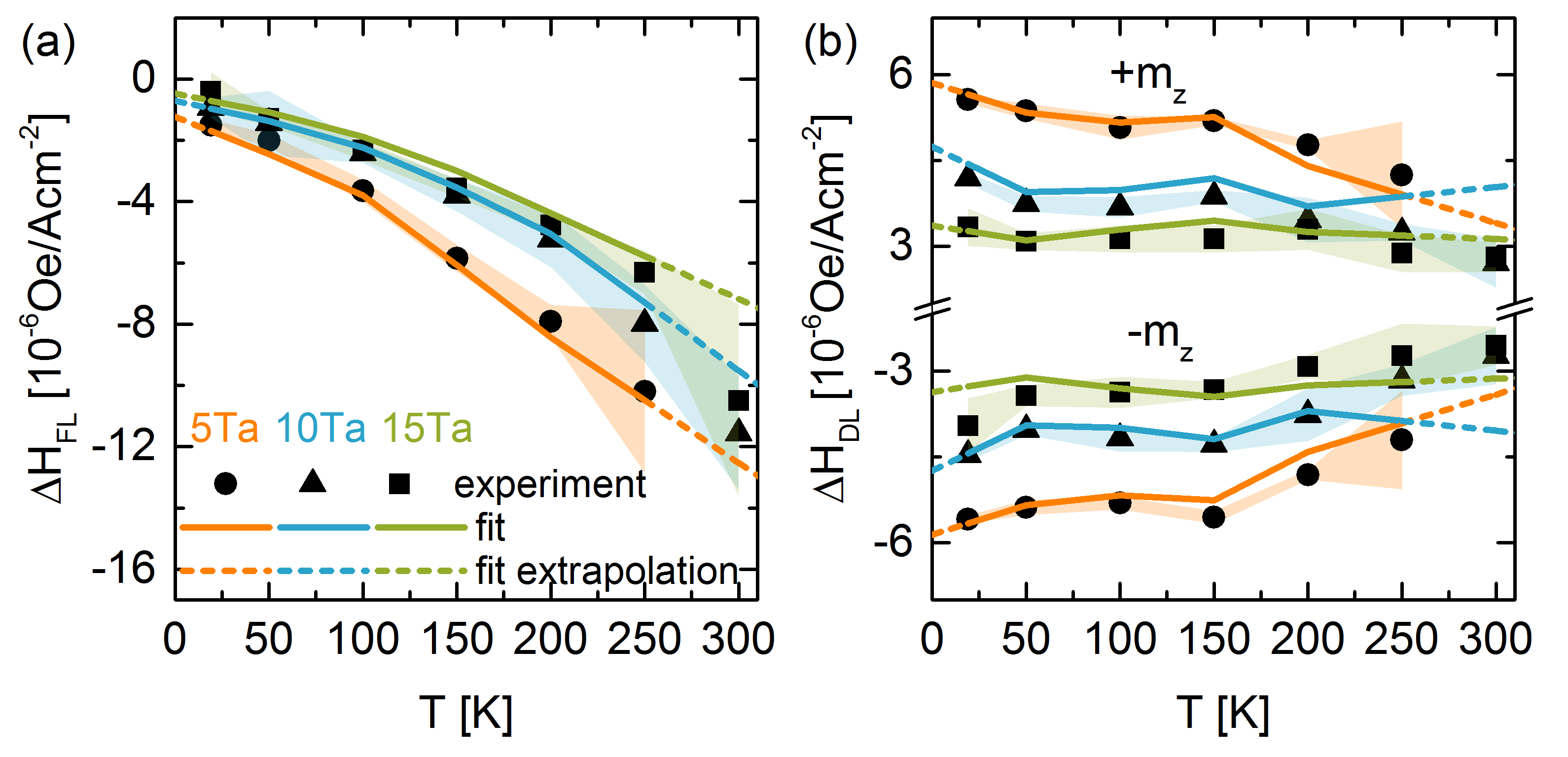}
\caption{Field-like $H_{FL}$ (a) and damping-like $H_{DL}$ (b) components of the torque-inducing effective magnetic field. Damping-like components have been shown for two indicated orientations of the CoFeB layer magnetization. The points correspond to experimental data while the solid lines represent the numerical fits. Shaded areas denote experimental error bars.}
\label{fig:Fig6}
\end{figure}

Temperature variation of the effective fields is shown in Fig.\ref{fig:Fig6}. Below 150 K, the longitudinal effective field, referred to as the damping-like (DL) field, is maintained at a constant level, and then in all three samples it decreases per module with increasing temperature (see Fig. \ref{fig:Fig6}(b)). The transverse effective field, referred to as field-like (FL) term, steadily decreases with increasing temperature. One can also note, that from room temperature down to 150 K, the field  $\Delta H_\mathrm{FL}$ dominates.
Both fields, $\Delta H_\mathrm{DL}$ and $\Delta H_\mathrm{FL}$, decrease with increasing Ta thickness, contrary to the results reported by Kim {\it et al}~\cite{kim2014anomalous}. However, it should be noted that our results cover a different range of Ta thickness. For thinner Ta layers the spin diffusion length has a decisive influence, while for thicker layers crystallographic structure plays a major role.

\begin{figure}[!ht]
\centering
\includegraphics[width=4.0 in]{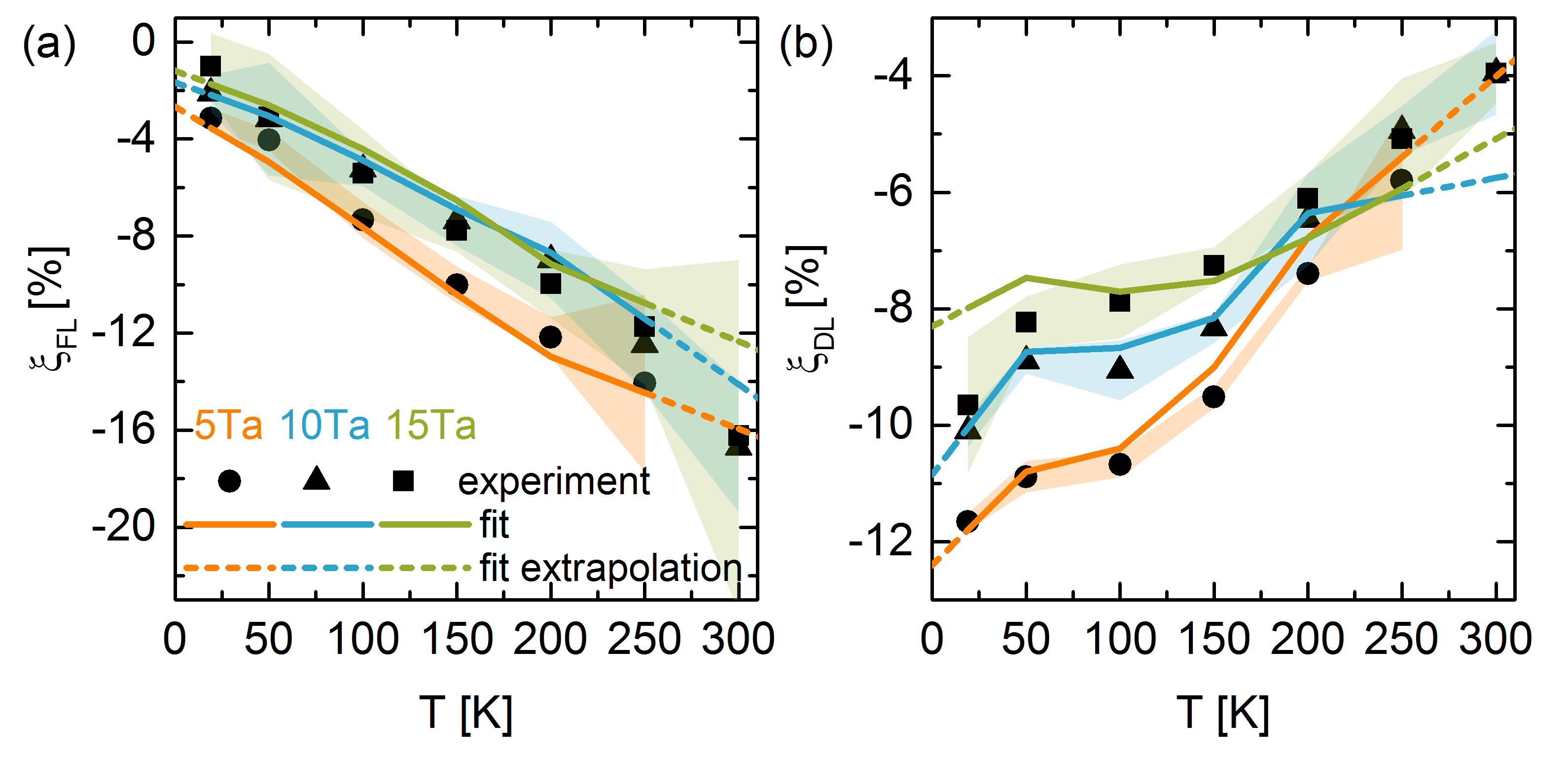}
\caption{Spin-torque efficiencies corresponding to the field-like  (a) and damping-like (b) components of the effective magnetic field. Shaded areas denote the experimental error bars. }
\label{fig:Fig7}
\end{figure}

Taking into account the current density in Ta and magnetic moment of CoFeB, the longitudinal (damping-like) and transverse (field-like) torque efficiences $\xi_\mathrm{FL(DL)}$ have be obtained from the formula $\xi_{FL(DL)}=(2|e|/\hbar )(\mu_0M_sd_F/J_N)\Delta H_{FL(DL)}$. Temperature variation of the evaluated spin torque efficiencies are presented in Fig.~\ref{fig:Fig7}. Discussion of the physical meaning of these quantities will be described in the next section.

%






\subsection*{Theoretical description}
   
To model spin transport and spin torques  we  consider interface as a distinct layer with its own properties such as spin diffusion length and spin Hall angle. Similar approach has been already used in the case of Pt/Py structures~\cite{pai2015dependence, wang2016giant}.
One of the key issues is  determination of an effective spin Hall angle of the structure, which generally can be a certain function of the spin Hall angle $\theta_{SH}^{N}$  of the Ta layer, referred to as a nonmagnetic (N) layer, and of the spin Hall angle $\theta_{SH}^{I}$ of the interface (I) layer, i.e.
$\Theta_{SH} = \Theta_{SH}(\theta_{SH}^{N},\theta_{SH}^{I})$. In general, the anomalous Hall effect can also play a role in conversion of charge current to spin current in N/F bilayer systems. In our case, however, this effect is small, as shown above.

A simple drift-diffusion equation for spin current in N/I/F (nonmagnetic/interface/Ferromagnetic) structure contains diffusion term resulting from spin accumulation at the interfaces and drift term due to SHE.
This equation written down for the nonmagnetic (N) and interface (I) layers take the form~\cite{chen2013theory,chen2016theory, kim2014anomalous, pai2015dependence}:
\begin{gather}
\mathbf{j}_{s}^{N}(z)=-\frac{1}{2e\rho_N}\frac{\partial \boldsymbol\mu_{s}^N(z)}{\partial z}-\theta_{SH}^{N}J_N\hat{\boldsymbol y}\,, \nonumber \\
\mathbf{j}_s^{I}(z)=-\frac{1}{2e\rho_I}\frac{\partial \boldsymbol\mu_{s}^I(z)}{\partial z}-\theta_{SH}^{I}J_I\hat{\boldsymbol y}\,,
\end{gather}
where $\rho_{N}$ and $\rho_{I}$ denote resistivity of the N and I layers, respectively, $e$ is the electron charge ($e<0$), $\boldsymbol\mu_{s}=\boldsymbol\mu_{\uparrow}-\boldsymbol\mu_{\downarrow}$ is the spin accumulation,  $\boldsymbol\sigma$ is the spin polarization of the SHE-induced spin current, while $\mathbf{J}_N$ and $\mathbf{J}_I$ denote the charge current density in the N and I layers.
In the following we assume $\theta_{SH}^{I}=\alpha\theta_{SH}^{N}$, with the proportionality coefficient $\alpha$ being constant.

Boundary conditions necessary for derivation of the spin accumulation $\boldsymbol \mu_s^{I}$ (and thus also spin current) take the form:
\begin{gather}
\mathbf{j}_{s}^{N}(z=d_I)=\mathbf{j}_{s}^{I}(z=d_I)\,, \nonumber \\
\mathbf{j}_{s}^{I}(z=0)=\mathbf{j}_s^{F|I}\,, \nonumber \\
\mathbf{j}_{s}^{N}(z=d_I+d_N)=0\,, \nonumber \\
\boldsymbol\mu_s^{I}(z=d_I)=\boldsymbol\mu_s^{N}(z=d_N)\,,
\end{gather}
where $d_{I,N}$ is the thickness of the I and N layers, respectively, and $z=0$ corresponds to the I/F interface.
Furthermore, $\mathbf{j}_s^{F|I}$ is the spin current at the F/I interface, taken on the interface layer side. This spin current obeys the boundary condition ~\cite{brataas2006non}:
\begin{equation}
e\mathbf{j}_s^{F|I}=G_r\hat{\mathbf{m}}\times\hat{\mathbf{m}}\times\boldsymbol\mu_s^{I}(0)+G_i\hat{\mathbf{m}}\times\boldsymbol\mu_s^{I}(0)\,,
\end{equation}
where $\hat{m}$ is a unit vector along magnetic moment of the F layer, $G_r\equiv \textrm{Re}\left[G_{mix}\right]$, $G_i\equiv \textrm{Im}\left[G_{mix}\right]$, and $G_{mix}$ is the so-called spin-mixing conductance.
The above equation is appropriate when the spin polarization of SHE-induced spin current is orthogonal to the magnetization of the magnet. If spin current is normal to $\hat{m}$, otherwise the spin current component parallel to $\hat{\mathbf{m}}$ may flow into the magnet and can induce anomalous Hall effect.

The spin current $\mathbf{j}_s^{F|I}$ creates a torque in the ferromagnetic layer, which can be expressed in terms on an effective field $\Delta\mathbf{H}$ as,
\begin{equation}
\frac{\partial \hat{\mathbf{m}} }{\partial t}=\boldsymbol\tau=\gamma\hat{\mathbf{m}}\times \Delta\mathbf{H}\,,
\end{equation}
where $\gamma$ is the gyromagnetic ratio and $\Delta\mathbf{H}$ is related to the spin current $\mathbf{j}_s^{F|I}$ via the formula
\begin{equation}
\Delta\mathbf{H}=\frac{\hbar}{2e}\frac{1}{\mu_0M_sd_F}\hat{\mathbf{m}}\times{\mathbf{j}_s^{F|I}}\,.
\end{equation}
The damping-like, $\Delta\mathbf{H}\cdot\hat{\mathbf{x}}\equiv \Delta H_{DL}$, and field-like, $\Delta\mathbf{H}\cdot\hat{\mathbf{y}}\equiv \Delta H_{FL}$ components of this effective field are
\begin{gather}
\Delta H_{DL}=\frac{\hbar}{2e}\frac{J_N}{\mu_0M_sd_F}\theta_{SH}^{N}\frac{\mathrm{tanh}
(\frac{d_N}{2\lambda_N})\mathrm{csch}(\frac{d_I}{\lambda_I})+\alpha\left[\mathrm{tanh}(\frac{d_I}{2\lambda_I})\mathrm{coth}(\frac{d_N}{\lambda_N})- \frac{\rho_I\lambda_I}{\rho_N\lambda_N}\right]}{\mathrm{coth}(\frac{d_N}{\lambda_N})\mathrm{coth}(\frac{d_I}{\lambda_I})+\frac{\rho_I\lambda_I}{\rho_N\lambda_N}}\frac{g_r\left(1+g_r \right)+g_i^2}{\left(1+g_r \right)^2+g_i^2}(-m_z) \,, \nonumber \\
\Delta H_{FL}=-\frac{\hbar}{2e}\frac{J_N}{\mu_0M_sd_F}
\theta_{SH}^{N}\frac{\mathrm{tanh}(\frac{d_N}{2\lambda_N})\mathrm{csch}(\frac{d_I}{\lambda_I})+\alpha\left[\mathrm{tanh}(\frac{d_I}{2\lambda_I})\mathrm{coth}(\frac{d_N}{\lambda_N})- \frac{\rho_I\lambda_I}{\rho_N\lambda_N}\right]}{\mathrm{coth}(\frac{d_N}{\lambda_N})\mathrm{coth}(\frac{d_I}{\lambda_I})+\frac{\rho_I\lambda_I}{\rho_N\lambda_N}}\frac{g_i}{\left(1+g_r \right)^2+g_i^2}\,,
\end{gather}
where $m_z=\pm 1$ is the projection of ferromagnet's magnetization onto the $z$ axis and $g_{r,i}$ defined as:
\begin{equation}
g_{r,i}=2G_{r,i}\frac{\mathrm{coth}(\frac{d_N}{\lambda_N})\mathrm{coth}(\frac{d_I}{\lambda_I})+\frac{\rho_I\lambda_I}{\rho_N\lambda_N}}{\frac{1}{\rho_I\lambda_I}\mathrm{coth}(\frac{d_N}{\lambda_N})+\frac{1}{\rho_N\lambda_N}\mathrm{coth}(\frac{d_I}{\lambda_I})}\,.
\end{equation}

\section*{Fitting to experiment and discussion}

In order to fit the above model to experimental data  we need to make some assumptions. First of all, we assume a constant spin diffusion length, $\lambda_N$, in the nonmagnetic layer (excluding the interface). In literature this parameter ranges from $\sim$1 nm to more than 3 nm~\cite{morota2011indication,liu2012spin}. In general, this parameter can also vary with temperature, however we neglect this variation due to possible compensation by changes in resistivity $\rho_N$. Moreover, both real, $G_r$, and imaginary, $G_i$, parts of spin-mixing conductance have been fixed by fitting to the data for a range of spin diffusion lengths and spin Hall angles. This fitting has shown that roughly $G_r(T)\sim\textrm{const}$ and $G_i(T)\sim T$.
This is consistent with the mixing counductance estimated for bulk Ta by \textit{ab initio} methods,~\cite{zwierzycki2005first,brataas2006non}
where, however, a crystalline phase was assumed and  the strong spin-orbit coupling was not taken into account.
The model has been then fitted to the experimental data for the available range of temperatures. Furthermore, the so-called spin memory loss (SML) parameter, defined as $\textrm{SML}=[1-\exp (-d_I/\lambda_I)]\cdot 100\%$, has been introduced. In numerical calculations we assumed $d_I/\lambda_I=0.05$, which corresponds to $\textrm{SML}\approx 4.9\%$.
This parameter has been assumed constant with respect to temperature. Such an assumption, however, may not hold at higher temperatures.
\begin{figure}[!ht]
\centering
\includegraphics[width=3in]{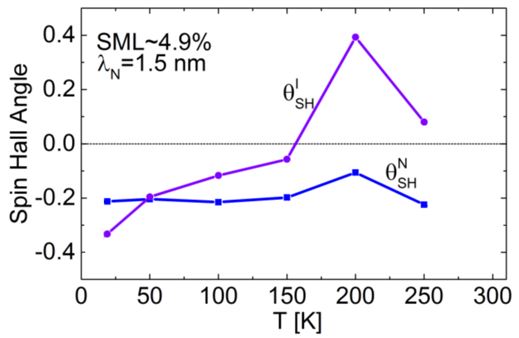}
\caption{Spin Hall angle of the interface and nonmagnetic metal Ta layers obtained from fits to experimental data with indicated parameters of spin memory loss (SML) and $\lambda_N$.}
\label{fig:Fig8}
\end{figure}

Figures~\ref{fig:Fig6} and \ref{fig:Fig7} show the best fit of the model to the experimental data on the field-like and damping-like components of the effective magnetic field and to the corresponding spin torque efficiencies. The absolute value of the field-like component of the spin-torque efficiency increases with increasing temperature, while the damping-like component decreases. This behavior of the field-like component can be explained by a dominant contribution of the imaginary part of the spin-mixing conductance. In the case of the damping-like component temperature dependence of $g_i$ cancels out and the dominating contribution comes from temperature dependence of the effective spin Hall angle.

Figure~\ref{fig:Fig8} shows the temperature dependence of the spin Hall angle in the nonmagnetic and interface layers, and the results indicate a strong interfacial effect. As the spin Hall angle in the nonmagnetic layer is roughly constant with respect to temperature, the interfacial spin Hall angle changes its sign for temperatures between 150-250K. This sign reversal indicates that the interfacial spin Hall angle should be treated as an effective spin current conversion coefficient, which is influenced by both nonmagnetic and ferromagnetic layers, and whose behavior at higher temperatures may differ from the behavior of nonmagnetic metals. Additional processes may play a role in this temperature range, resulting  in higher spin memory loss. Opposite signs of the spin Hall angles in the nonmagnetic and interfacial layers have occurred also in the analysis of spin pumping induced ISHE in Bi/Py bilayer system, where scattering on impurities has been considered as a possible explanation of this behavior \cite{hou2012interface}. The second interesting feature of  the interface spin Hall angle is its magnitude, which in the vicinity of 200 K is remarkably larger (though it has opposite sign) than that in the Ta layer. This, however, is reasonable as the charge current density in the interface layer may be smaller than that in the Ta layer. Apart from this,  additional extrinsic mechanism of SHE in the interface layer can occur due to intermixing (scattering on magnetic impurities). We also note, that the interfacial spin Hall angle in a Pt/Py structure has been estimated to be 25 times larger than the spin Hall angle in a single Pt layer \cite{wang2016giant}, whereas the interfacial contribution in a Bi/Py bilayer  has been estimated to be ca. 4 times larger than the contribution from a single Pt layer \cite{hou2012interface}.

In conclusion, we have investigated the spin Hall effect and spin torques in Ta/CoFeB structures. We have shown that strong intermixing at the interface leads to a magnetically dead layer, which is especially thick for structures with 5 nm Ta layer. To account for the experimentally determined damping-like and field-like effective fields, we applied a drift-diffusion model assuming the interfacial region as a distinct layer. By fitting the theoretical model to experimental data we have determined the spin Hall angle in the Ta layer and in the Ta/CoFeB interface layer. The interface spin Hall angle was shown to change its sign with increasing temperature. Moreover, at temperatures around 200 K its magnitude is larger than that of spin Hall angle in the tantalum layer.

\section*{Methods}
 All samples were deposited using magnetron sputtering in a Singulus Timaris PVD Cluster Tool System on thermally oxidized Si(001) 4-inch wafers. The Ta and CoFeB layers were magnetron sputtered at an argon pressure of $2.7\times10^{-3}$mbar. For CoFeB, a linear dynamic deposition (LDD) wedge technology was used to achieve a smooth gradation of film thickness. Samples with different $t_{FM}$ layer were used to study the formation of a magnetic dead layer. However, for the SHE measurements all structures had a constant CoFeB thickness, $t_{FM}$ = 0.91 nm. For this thickness, the as-deposited samples exhibited an uniaxial in-plane anisotropy, which turned into perpendicular anisotropy after 20 minutes post-deposition annealing at 330$^\circ$.

Continuous layers were used for structural and magnetic characterization. $\theta$$-2$$\theta$ X-ray Diffraction was employed to study the crystallograhic structure. Interface roughness was examined by X-ray Reflectivity and the surface morphology of the Ta layers was measured using Atomic Force Microscopy. High-resolution structural characterization of multilayer samples with 5 nm and 15 nm Ta was carried out using a JEOL 2200FS Transmission Electron Microscope with double Cs correctors. Magnetic properties of the FM layer were probed using a LakeShore 7407 vibrating sample magnetometer (VSM) with an LN2 cryostat. The change of magnetization was measured in a magnetic field ($\pm$ 50 mT) at temperatures ranging from 80 K to 300 K. Using e-beam lithography and ion-beam etching methods, the samples were patterned into 10 $\mu$m width and 40 $\mu$m long Hall bars with 100 $\times$ 100 $\mu$m$^2$ contact pads. Additional structures were prepared for 4-point resistance measurements.

\section*{Acknowledgements}

M.C. and Ł.K. received financial support from the NANOSPIN under Grant PSPB-045/2010 from Switzerland through the Swiss Contribution, W.S. acknowledges National Science Center, Poland, Grant No. 2015/17/D/ST3/00500, J.K. acknowledges DEC-2012/05/E/ST7/00240, T.S. acknowledges Grant Harmonia No. UMO-2012/04/M/ST7/00799. S.v.D. and L.Y. acknowledge financial support from the Academy of Finland (Grant Nos. 12286361 and  13293929) and the European Research Council (ERC-2012-StG 307502). Microfabrication process was performed at Academic Center for Materials and Nanotechnology of AGH University. TEM analysis was conducted at the Aalto University OtaNano - Nanomicroscopy Center (Aalto-NMC). M.C. acknowledges AGH University Dean’s Grant No. 15.11.230.269. Ł.K. would like to thank Anna Dyrda\l{} for helpful discussions.

%
%
%
%
%

\end{document}